\documentclass[epj]{svjour}
\usepackage{graphics}
\begin{document}
\title{Global Fits of the CKM Matrix}
\author{G. Eigen\inst{1}  G.P. Dubois-Felsmann\inst{2}, D.G. Hitlin\inst{2}, \and
F.C. Porter\inst{2}
}                     
%
%
\institute{Dept. of Physics, University of Bergen \and 
Lauritsen Laboratory, Caltech}
%
%
\abstract{We report upon the present status of global 
fits to Cabibbo-Kobayashi-Maskawa matrix.}
%
%
\maketitle
\vskip -1.5cm
\section{Introduction}
\label{intro}
The three-family Cabibbo-Kobayashi-Maskawa (CKM) \break quark-mixing matrix 
is a key element of the Standard Model (SM). The nine complex CKM elements
are completely specified by three mixing angles and one phase that is 
responsible for $C\!P$ violation in the SM. Measuring the CKM matrix elements
in various ways provides consistency tests of the matrix elements itself 
and with unitarity. Any significant inconsistency with the SM would indicate 
the presence of new physics.

A convenient parameterization of the CKM matrix is the Wolfenstein
approximation \cite{bib:wolfenstein}, which to order $O(\lambda^3)$ is given by:

\begin{equation}
\label{eqn:CKMmatrix}
    V = \pmatrix{1-{\lambda^2\over 2}  &
         \lambda & A\lambda^3(\rho-i\eta) \cr
         -\lambda &
           1-{\lambda^2\over 2} & A\lambda^2 \cr
         A\lambda^3(1-\bar\rho-i\bar\eta) & -A\lambda^2
 & 1 \cr} +O(\lambda^4),
\end{equation}

\noindent
where $\lambda=0.2241\pm 0.0033$ is the best-known parameter measured in 
semileptonic $K$ decays, $A =0.82$ is determined from semileptonic 
$B$ decays to charmed particles with an accuracy of $\simeq 6\%$ and  
$\bar \rho = \rho \cdot (1-\lambda^2 /2)$ and 
$\bar \eta = \eta \cdot (1-\lambda^2 /2)$ are least-known.

The unitarity of the CKM matrix yields six triangular relations of which 
$V_{ud}V^*_{ub}+V_{cd}V^*_{cb}+V_{td}V^*_{tb}=0$ is well-suited for
experimental tests. In order to determine the apex of the
unitarity triangle $(\bar \rho, \bar \eta)$ presently 
eight measurements are used as input, the $B$ semileptonic branching 
fractions ${\cal B}(B\rightarrow X_c \ell \nu)$, 
${\cal B}(B\rightarrow X_u \ell \nu)$, and 
${\cal B}(B\rightarrow \rho \ell \nu)$,
the normalized $B\rightarrow D^* \ell \nu$ rate at zero recoil, 
${\cal F}(1)|V_{ub}|^2$, the
$B^0_d$ and $B^0_s$ oscillations frequencies $\Delta m_{B_d}$ and $\Delta 
m_{B_s}$, the parameter $|\epsilon_K|$ that specifies  $C\! P$ violation in 
the $K^0 \bar K^0$ system, as well as $\sin 2 \beta$ which is measured in 
$C\! P$ asymmetries of charmonium $K^0_S \ (K^0_L)$ final states. Though 
many of these measurements themselves are rather precise their translation 
to the $\bar \rho- \bar \eta $ plane is affected by large non-gaussian 
theoretical uncertainties. Various approaches,  which treat theoretical 
errors in different ways, can be found in the literature  
[2,3,4,5,6].

\section{The Scan Method}
The scan method is an unbiased procedure for extracting 
$A, \bar \rho, \bar \eta$ from measurements. We select observables that 
allow us to factorize their predictions in terms of 
theoretical quantities $T_i$ that have an {\it a priori} unknown (and
likely non-gaussian) error distribution 
$(\Delta_i)$, other observables, and the CKM dependence
expressed as functions of Wolfenstein parameters. As an example,
consider the charmless semileptonic branching fraction 
for $B \rightarrow \rho \ell \nu$, which is predicted to be 
${\cal B}(B\to\rho\ell\nu) = |V_{ub}|^2 \cdot \widetilde \Gamma_{\rho\ell\nu} 
\cdot \tau_{B}$, where $\tau_{B^0}$ is the $B^0$ lifetime and 
$\widetilde \Gamma_{\rho\ell\nu}$ is the reduced rate affected by 
non-gaussian uncertainties. This analysis treats
eleven theoretical parameters with non-gaussian errors,
the reduced inclusive semileptonic rates $ \widetilde \Gamma_{X_u\ell\nu}$ and
$ \widetilde \Gamma_{X_c\ell\nu}$, the form factor for 
$B \rightarrow D^* \ell \nu$ at zero recoil, ${\cal F}_{D^*}(1)$, the
bag factors of the $K^0$ and $B^0$ systems, $B_K$ and$B_B$, the $B^0$ decay 
constant $f_B$, $\xi^2 = f^2_{B_s}/f^2_{B_d} B_{B_s}/B_{B_d}$ and the QCD 
parameters $\eta_1, \eta_2, \eta_3$ and $\eta_B$. 

We perform a $\chi^2$ minimization based on a frequentist approach by
selecting a specific value for each $T_i$ within the allowed range 
(called a model). We perform individual fits for many models scanning over 
the allowed non-gaussian ranges of the $T_i$ parameter space.
The QCD parameters are not scanned; their small errors are treated 
in the $\chi^2$ as gaussian. For theoretical quantities calculated on the
lattice, which have gaussian errors ($B_K$, $B_B$, $f_B$ and $\xi$)
we add specific $\chi^2$ terms. To account for correlations between 
observables that occur in more than one prediction, such as the masses of 
the $t$-quark, $c$-quark, and $W$-boson, $B$ hadron lifetimes, $B$ hadron 
production fractions and $\lambda$, we include additional terms in the 
$\chi^2$ function. 

We consider a model to be consistent with the data if the fit probability 
yields $P(\chi^2) > 5\%$. We determine the best estimate for each 
of the 17 fit parameters and plot a $95\%$ confidence level (C.L.) contour in 
the $\bar \rho- \bar \eta$ plane. We overlay the $\bar \rho- \bar \eta$ 
contours of all accepted fits. In order to study 
correlations among the $T_i$ and constraints the data impose we perform
global fits with non-gaussian theory errors scanned over a 
$\pm 5 \Delta$ wide range (see section 5).

\subsection{Treatment of $\Delta m_{B_s}$}
Since $B^0_s \bar B^0_s$ oscillations have not been observed yet, a lower 
limit on $\Delta m_{B_S}$ at 95\%~C.L. has been determined 
by combining analyses of different experiments using the amplitude 
method~\cite{bib:moser}. To incorporate $\Delta m_{B_S}$ into the $\chi^2$ 
function, we use a new approach that is based upon the significance of a 
$\Delta m_{B_s} $ measurement~\cite{bib:cern}:
\begin{equation}
 S = \sqrt{{N\over 2}} f_{B_s} (1-2w) e^{-{1\over 2}(\Delta m_s \sigma_t)^2},
\end{equation}
where $N$ is the sample size, $f_{B_s}$ is the $B_s$ purity, $w$ is the 
mistag fraction, and $\sigma_t$ is the resolution. Substituting $C$ for 
$\sqrt{{N\over 2}} f_{B_s} (1-2w)$ and interpreting $S$ as the number of 
standard deviations by which $\Delta m_{B_s}$ differs from zero, 
$S=\Delta m_{B_s}/\sigma_{\Delta {m_{B_s}}}$, we may define a contribution 
to the $\chi^2$ from the $\Delta m_{B_s}$ measurements as:
\begin{equation}
 \chi^2_{\Delta {m_{B_s}}} = C^2\left(1-{\Delta\over\Delta {m_{B_s}}}\right)^2
   e^{-(\Delta {m_{B_s}}\sigma_t)^2},
\end{equation}
where $\Delta$ is the best estimate according to experiment. The values of 
$(\Delta, C^2, \sigma_t)$ are chosen to give a minimum at $17~ \rm ps^{-1}$, 
and a $P(\chi^2)=5\%$ at $\Delta {m_{B_s}}=14.4~\rm ps^{-1}$. 
In the region of small $\chi^2$, this function exhibits similar general 
features as that used in our previous global fits~\cite{bib:eps}, while
it does not suffer from numerical instabilities arising from multiple minima.
The two functions deviate at large values of $\chi^2$, where in any case
poor fits result. 


\begin{table}
\caption{Measurement inputs used in $\chi^2$ minimization}
\label{tab:expin}       
\begin{tabular}{lll}
\hline\noalign{\smallskip}
Observable & Value & Comment  \\
\noalign{\smallskip}\hline\noalign{\smallskip}
${\cal B}(B \rightarrow X_u\ell\nu) $&$(2.03\pm0.22_{exp}\pm0.31_{th})\times10^{-3} $&  $\Upsilon(4S)$ \\
${\cal B}(B\rightarrow X_u\ell\nu) $&$ (1.71\pm0.48_{exp}\pm0.21_{th})\times10^{-3}$&  LEP \\
${\cal B}(B\rightarrow X_c\ell\nu)$ &$ 0.1070\pm 0.0028$& $\Upsilon(4S)$ \\
${\cal B}(B\rightarrow X_c\ell\nu)$ &$ 0.1042\pm 0.0026 $&    LEP \\
${\cal B}(B\rightarrow \rho\ell\nu)$& $ (2.68\pm0.43_{exp}\pm0.5_{th})\times10^{-3}$&  
CLEO/{\it BABAR} \\
$ |V_{cb} |F(1)$&$0.0388\pm 0.005 \pm 0.009$&           LEP/CLEO/Belle \\
$  \Delta m_{B_d}$& $ (0.503\pm 0.006) \rm  ps^{-1} $&         world average \\
$ \Delta m_{B_s}$ & $ > 14.4~ \rm ps^{-1}  @95\%C.L.$        &        LEP\\
$ |\epsilon_K |$&$ (2.271\pm 0.017)\times 10^{-3}$ & PDG 2000 \cite{bib:groom}     \\
$  \sin 2 \beta$& $ 0.731\pm 0.055$ &             
{\it BABAR}/Belle \\
 $  \lambda$&$ 0.2241\pm 0.0033$ &        world average \\
\noalign{\smallskip}\hline
\end{tabular}
\end{table}

\begin{figure}
\hskip 1.0cm
\resizebox{0.45\textwidth}{!}{\includegraphics{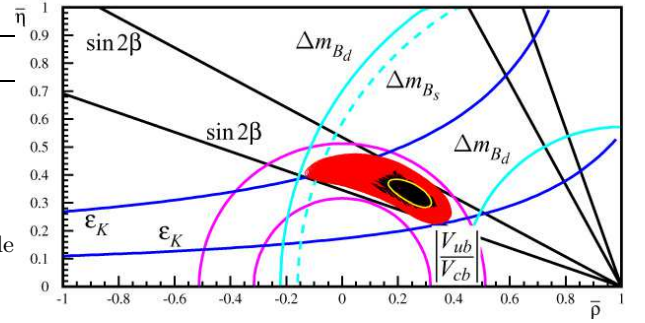}}
\caption{Results of the global fit in the $\bar \rho- \bar \eta$ plane.}
\label{fig:com}       
\end{figure}

\begin{figure}
\hskip 1.0cm
\resizebox{0.42\textwidth}{!}{\includegraphics{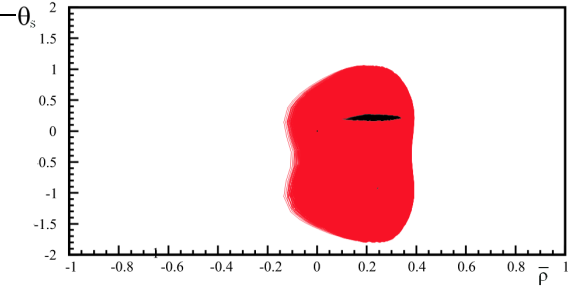}}
\caption{Fit results in $\theta_s- \bar \rho$ plane from 
$a_{\phi K^0_S}$.}
\label{fig:phiks}       
\end{figure}

\begin{table}
\caption{Theoretical parameter with non-gaussian errors}
\label{tab:thin}       
\begin{tabular}{ll}
\noalign{\smallskip}\hline\noalign{\smallskip}
$0.87 \leq {\cal F}_{D^*}(1) \leq 0.95 $ & 
$38.0 \leq \widetilde \Gamma(c\ell\nu) \leq 41.5~ \rm ps^{-1} $ \\ 
$12.0 \leq  \widetilde \Gamma(\rho \ell\nu) \leq 22.2~ \rm ps^{-1} $& 
$54.8 \leq \widetilde \Gamma(u\ell\nu) \leq 79.6~ \rm ps^{-1} $ \\
\hline
$0.72 \leq B_K \leq 1.0$ & $\sigma_{B_K} = 0.06$ (gaussian) \\
$211 \leq f_{B_d} \sqrt{B_{B_d}} \leq 235~ \rm MeV $& 
$\sigma_{f_B\sqrt{B_B}} = 33~\rm MeV$ (gaussian) \\ 
$1.18 \leq \xi \leq 1.30 $ & $\sigma_\xi=0.04$ (gaussian) \\
\hline
$0.54 \leq \eta_B \leq 0.56$ & 
$1.0 \leq \eta_1 \leq 1.64$ \\
$0.564 \leq \eta_2 \leq 0.584 $ & 
$0.43 \leq \eta_3 \leq 0.51$ \\
\noalign{\smallskip}\hline
\end{tabular}
\end{table}

\begin{table}
\caption{Results of $95\%$ C.L. range for $\bar \rho, \bar \eta, \alpha$ 
and $\gamma$ from the global fits shown in figure~\ref{fig:com}. For
comparison results from RFIT and the Bayesian method are also given.}
\label{tab:result}       
\begin{tabular}{llll}
\hline
parameter &Scan method & RFIT \cite{bib:cern} & Bayesian \cite{bib:cern} \\
\hline\noalign{\smallskip}
$\bar \rho$ & -0.13 to 0.40 & 0.091 to 0.317 & 0.137-0.295 \\
$\bar \eta$ & 0.22 to 0.48  & 0.273 to 0.408 & 0.295-0.409 \\
$\alpha$   & $50.4^0$ to $126.6^0$   &  &  \\
$\gamma$   & $34.4^0$ to $91.7^0 $  & $42.1^0$ to $75.7^0$   & $47.0^0$ to 
$70.0^0$ \\
\noalign{\smallskip}\hline
\end{tabular}
\end{table}
\section{Results of the global Fit}
Figure~\ref{fig:com} shows the result of scanning all $T_i$ simultaneously 
within $\pm 1 \Delta$ of their allowed range except for the QCD parameters.
We have used the input measurements summarized in table~\ref{tab:expin} and  
ranges for the $T_i$ listed in table~\ref{tab:thin}. The black points 
represent the best estimates of $(\bar \rho, \bar \eta)$ for each model 
that is consistent with the data. The grey region shows the overlay of all 
corresponding $95 \%\ \rm C.L.$ $\bar \rho-\bar \eta$ contours. 
For reference, the light ellipse
depicts a typical contour. To guide the eye the $95\% \ \rm C.L.$ bounds on 
$|V_{ub}/V_{cb}|$, $|\epsilon_K|$, $\Delta m_{B_d}$ and $\sin 2 \beta$ as 
well as the lower bound on $\Delta m_{B_s}$ are also plotted. From 
these fits we derive $95\% \ \rm C.L.$ ranges for $\bar \rho, \bar \eta, 
\alpha$ and $\gamma$ that are listed in table~\ref{tab:result}. For 
comparison, recent results from two other global fits (RFIT 
\cite{bib:ckmfitter}, Bayesian fit \cite{bib:ciuchini}) are also shown.

Using the same source of inputs, several 
differences exist between the scan method and the other two approaches. 
First, we scan separately over the inputs of exclusive and inclusive 
$b\rightarrow u \ell \nu$ and $b \rightarrow c \ell \nu$ measurements. 
Second, we use a different approach to incorporate 
$\Delta m_{B_s}$. While in the Bayesian method theoretical quantities are
parameterized in terms of gaussian and uniform distributions, we make 
no assumptions about their shape. Thus, the Bayesian fits tend to
produce a smaller region in the $\bar \rho- \bar \eta$ plane and are
more sensitive to fluctuations than corresponding fits in the scan method.
In RFIT, the  $\bar \rho- \bar \eta$ plane is scanned to find a solution in 
the theoretical parameter space. Since in RFIT a central region with equal 
likelihood is determined, it is not possible to give
probabilities for individual points. 
In contrast, in the scan method 
individual contours have a statistical meaning, with the center point
yielding the highest probability. Since the mapping of the 
theory parameters to the $\bar \rho- \bar \eta$ plane is not 
one-to-one, it is possible in the scan method to track which values of  
$(\bar \rho, \bar \eta)$ are preferred by the theory parameters.

\section{Search for New Physics}
The decay $B \rightarrow \phi K^0_S$ that proceeds via a 
$b \rightarrow s \bar ss$ penguin loop is expected measure $\sin 2 \beta$ in 
the SM to within $\sim\!4\%$. New physics contributions, however, may 
introduce a new phase $\theta_s$ that may change the $C \!  P$ asymmetry 
$a_{\phi K^0_S}$ significantly from $a_{J/\psi K^0_S}$. The 
{\it BABAR}/Belle
average of $S_{\phi K^0_S}=-0.39 \pm 0.41$ has been updated this summer  
yielding $S_{\phi K^0_S}=-0.14 \pm 0.33$~\cite{bib:hfag}. The deviation from 
$\sin 2 \beta$ has remained at $\sim 2.6 \sigma$. In our global fit we 
introduce a new phase $\theta_s$. Figure~\ref{fig:phiks} shows the
overlay of all resulting contours in the 
$\theta_s- \bar\rho$ plane that have acceptable fit probabilities. 
Presently, the phase is consistent with zero as expected in the SM.

Physics beyond the SM may affect $B^0_d \bar B^0_d$ mixing and $C \! P$ 
violation in $B \rightarrow J/\psi K^0_s$ and $B \rightarrow \pi \pi$. 
Using a model-independent analysis \cite{bib:nir} 
we can introduce a scale parameter, 
$r_d$, for $B^0_d \bar B^0_d$ mixing and an additional phase, $\theta_d$, 
for parameterizing $a_{\psi K^0_s}$. 
In the SM we expect $r_d=1$ and $\theta_d=0$.
With present uncertainties $r_d$ and $\theta_d$ are consistent with 
the SM expectations (see \cite{bib:eps}).

\begin{figure}
\hskip 1.2 cm
\resizebox{0.4\textwidth}{!}{%
  \includegraphics{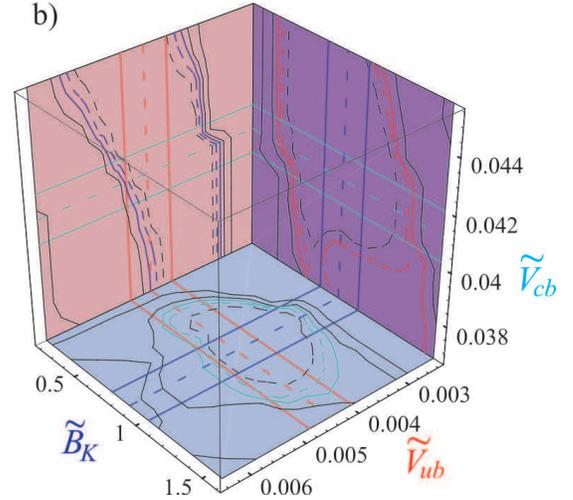}}
\caption{Contours of the theory parameters $\widetilde B_K- 
\widetilde V_{ub}-\widetilde V_{cb}$ both resulting from inclusive reduced 
semileptonic rates for fit probabilities $P(\chi^2)>32\%$ after scanning
$\widetilde B_K, \widetilde f_B\sqrt{B_B}, \widetilde \xi, 
\widetilde \Gamma(B \rightarrow X_c \ell \nu$ and
$\widetilde \Gamma(B \rightarrow X_u \ell \nu$ over $\pm 5 \Delta_i$ range.} 
\label{fig:in3}       
\end{figure}

\section{Visualizing the role of theoretical errors}
In addition to the global fits in the $\bar \rho- \bar \eta$ plane, we  
explore the impact of measurements on the theoretical parameters and 
their correlations. We typically scan theory parameters within 
$\pm 5 \Delta$ and denote them with $^\sim$. Presently, we use either 
exclusive or inclusive  $\widetilde V_{ub}, \widetilde V_{cb}$ information
and plot contours for three of the five scanned theoretical parameters for
different conditions. An example is shown in Figure~\ref{fig:in3}, where 
we have scanned inclusive $\widetilde V_{ub}$, inclusive
$\widetilde V_{cb}$, $\widetilde B_K$, 
$\widetilde {f_{B_d} \sqrt{B_{B_d}}}$ and $\widetilde \xi$. For 
$\widetilde V_{ub}$, $\widetilde V_{cb}$ and $\widetilde B_K$ we plot 
two-dimensional contours on the surface of a cube.
In each plane five contours are visible. The outermost contour (solid black) 
results from requiring a fit probability of $> 32 \%$. The next contour 
(also solid black) is obtained by restricting all other undisplayed 
theory parameters to their allowed range of $\pm 1 \Delta$. The third 
solid line results by fixing the parameter orthogonal to plane to the allowed 
range, while the outer dashed line is found if the latter parameter 
is fixed to its central value. The internal dashed black line is obtained 
by fixing all undisplayed parameters to their central values. 
Further details, other combination plots and results for 
exclusive $\widetilde V_{ub}$ and $\widetilde V_{cb}$ scans are discussed in 
\cite{bib:eps}.

\section{Conclusion}
The scan method provides a conservative, robust method that treats
non-gaussian theoretical uncertainties in an unbiased way. 
This reduces conflicts with the SM resulting from unwarranted assumptions
concerning the theoretical uncertainties, which is important in  
searches for new physics. The scan methods yields significantly larger ranges 
for the $\bar \rho- \bar \eta$ plane than the Bayesian method. Presently, all 
measurements are consistent with the SM expectation due to the large 
theoretical uncertainties. The deviation of $a_{\phi K^0_s}$ from 
$\sin 2 \beta$ measured in charmonium 
$K^0_S \ (K^0_L)$ modes is interesting but not yet significant. 
Model-independent parameterizations will become important in
the future when theory errors are further reduced. 
%
%

\end{document}